\documentclass[prl,twocolumn,showpacs,amsmath,amssymb,floatfix]{revtex4}
\usepackage{color,graphics,epsfig,rotating}
\usepackage{graphicx}% Include figure files
\usepackage{dcolumn}% Align table columns on decimal point
\usepackage{bm}% bold math

\begin{document}

%\title{Electron-phonon superconductivity in doped
%  solid picene}

%\title{Solid picene: a possible exotic electron-phonon superconductor}

\title{Vibrational spectrum and electron-phonon coupling of doped
  solid picene from first principles}

%Electron-phonon superconductivity in solid picene}
%the case of picene}

\author{Alaska Subedi}
\affiliation{Max Planck Institute for Solid State Research,
  Heisenbergstrasse 1, D-70569 Stuttgart}

\author{Lilia Boeri}
\affiliation{Max Planck Institute for Solid State Research,
  Heisenbergstrasse 1, D-70569 Stuttgart}

\date{\today}

\begin{abstract}
  We study superconductivity in doped solid picene (C$_{22}$H$_{14}$)
  with linear response calculations of the phonon spectrum and
  electron-phonon ($ep$) interaction. We show that the coupling of the
  high-energy C bond-stretching phonons to the $\pi$ molecular
  orbitals for a doping of $\sim$3 electrons per picene molecule is
  sufficiently strong to reproduce the experimental $T_c$ of 18 K
  within Migdal Eliashberg theory.  For hole doping, we predict a
  similar coupling leading to a maximum T$_c$ of 6 K. However, we
  argue that, due to its molecular nature, picene may belong to the
  same class of strongly correlated $ep$ superconductors as
  fullerides.  We propose several experimental tests for this
  hypothesis and suggest that intercalated hydrocarbons with different
  arrangements and numbers of benzene rings may be used to study the
  interplay between $ep$ interaction and strong electronic
  correlations in the highly non-adiabatic limit.
\end{abstract}


% Ab initio calculations (electronic structure of atoms and
% molecules), 31.15.A-

% Organic superconductors, 74.70.Kn

% Electron-phonon interactions
%    lattice dynamics, 63.20.kd

\maketitle

%\section{Introduction}

The field of superconductivity has witnessed several important
discoveries in the last ten years that, together with new synthesis
and manipulation techniques, have enormously advanced our
understanding of this phenomenon.
These include MgB$_2$ with $T_c$ of 39 K,\cite{naga01} boron-doped diamond with
$T_c$ as high as 11.5 K,\cite{ekim04} intercalated graphite\cite{YbC6:weller:syn} with a
$T_c$ of 14.5 under pressure,\cite{emer05} and iron based superconductors\cite{kami08} with
$T_c$ of up to 56 K.\cite{wang08}
However, in order to be able to design superconductors with desired
properties, additional efforts are  needed to understand and
control the material parameters relevant for superconductivity in
different classes of compounds.

Carbon based compounds offer many possibilities in this regard
because their structure can be easily manipulated 
at the atomic level and they
generally posses high phonon frequencies that have substantial
electron-phonon ($ep$) coupling.
Besides many conventional phonon-mediated superconductors, this class
also comprises alkali-intercalated fullerides (A$_3$C$_{60}$) with
$T_c$ of up to 40 K.\cite{heba91,pals95}  These
$\pi$-bonded
molecular solids have a rich phase diagram determined by a non-trivial
interplay between strong local $ep$ interactions and on-site Coulomb
correlations in a highly non-adiabatic regime.~\cite{gunn:book,mass09,baffo} 
The physics of this strongly correlated $ep$ superconductors
is of great fundamental interest. Therefore, identifying other compounds in 
this class is highly desirable.
For this reason, the discovery of superconductivity in K and Rb doped
picene with $T_c$ of up to 18 K by Mitsuhashi \textit{et al.}\ last
year\cite{mits10} is of great significance.
% as it provides an
% opportunity to increase our understanding of superconductivity near
% Mott insulating phase, superconductivity beyond Eliashberg theory...

% In this paper, we show that the recently discovered alkali-doped
% picene could be such an example, using linear response of the phonon
% spectra and $ep$ interaction in the rigid band
% approximation.
%
In this paper, we use linear response calculations of the phonon
spectra and $ep$ interaction~\cite{foot:calc}
 within the rigid band approximation (RBA) 
to show that alkali-doped picene could be another case
of strongly correlated  $ep$
superconductors.
Our results complement previous theoretical and experimental studies
of this compounds, which have focused on electronic properties and
correlations.\cite{kosu09, okaz10, roth10, andr10, kim10, giov10,
  roth11} We obtain a value of the $ep$ coupling constant that is
large enough to explain superconductivity with $T_c = 18$ K in
Migdal-Eliashberg theory. However, since the bandwidths of the
conduction and valence bands in solid picene are relatively small ($\sim 0.3$--1
eV), we argue that a local approach that includes both the $ep$
coupling and Coulomb correlations on an equal footing may be more
appropriate and also provide parameters that are relevant to these
type of models.
Finally, we discuss possible experimental tests that could confirm or
exclude the scenario we propose.

\begin{figure} %% [tbp]
  \includegraphics[width=\columnwidth]{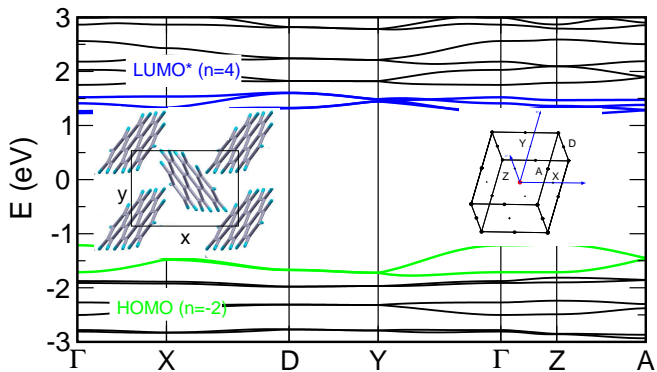}
  \caption{(Color online) Calculated LDA electronic structure of solid
    picene.  In green (light gray) and blue (dark gray) we show the
    two HOMO and the four LUMO$^*$ derived bands, respectively, which
    are accessible by hole or electron doping. $n$ is the maximum
    number of electrons per picene molecule that can be doped.  The
    two insets show the unit cell of solid picene in the $xy$ plane
    and its Brillouin Zone (BZ), respectively.}
  \label{fig:ele}
\end{figure}

Picene  (C$_{22}$H$_{14}$) is a polycyclic aromatic hydrocarbon formed by
five benzene rings juxtaposed in an armchair arrangement.
Solid picene is an insulator, with a gap of $\sim$3.3 eV, which
can form in either monoclinic or orthorhombic structure.\cite{okam08}
Since the superconducting samples of Ref.~\onlinecite{mits10} have a
monoclinic $P2_1$ structure, we employed experimental lattice
parameters for undoped picene in this variant with $a = 8.480, b =
6.154, c = 13.515$ \AA, $ \beta = 90.46^\circ$.\cite{de85}
We relaxed the internal coordinates so that every component of force
was less than $10^{-4}$ Ry/au. This changed the C-C and C-H bond
lengths by less than 2.5\% and was needed in order to have a
dynamically stable structure with real phonon
frequencies.~\cite{foot:freq}
The unit cell contains two picene molecules placed in a herringbone
arrangement in the $xy$ plane (shown in the inset of
Fig.~\ref{fig:ele}) which are then stacked along $z$ direction. The shortest
intermolecular C-C distance is 3.5 \AA\ within the $xy$ plane and 3.9 \AA\
along $z$ (i.e. along the length of the picene molecule).
These intermolecular distances are more than twice the typical intramolecular
nearest-neighbour C-C distance of 1.4 \AA.
In the doped superconducting crystals, the $a$ axis expands while the $b$ and
$c$ axes shrink. It is inferred from this that the dopants most likely
occupy intra-layer positions since inter-layer intercalation would
have caused the $c$ axis to expand. However, the detailed experimental
structure of doped picene is not available.
For this reason, we simulated the effect of doping using the rigid
band approximation (RBA) and did not include explicitely the dopants
in the calculations.  The role of the dopant, which should be minor,
is discussed in relation to graphite intercalated compounds (GIC).

The resulting electronic structure, which agrees well with previous
density functional theory
calculations,\cite{kosu09,roth10,andr10,giov10} is shown in the main
panel of Fig.~\ref{fig:ele}.  The zero of the energy in the figure is
chosen as the middle of the calculated $\sim$2.4 eV gap that separates
the the bonding valence bands from the anti-bonding conduction bands
in undoped picene.
All bands in the energy range shown in the figure have mostly C $p_z$
character. They are lumped into small subsets with very small in- and
out-of-plane dispersion reflecting the molecular nature of the solid.
This is similar to the electronic structure of fullerenes whose
conduction bands also have C $p_z$ character and are very
narrow.\cite{gunn:book}
In superconducting picene (K/Rb$_{3}$C$_{22}$H$_{14}$) the $\sim$3
electrons donated from the alkali atoms to each molecule populate the
4 bands immediately above the gap. The Fermi surface (not shown) is
large without any significant two-dimensional nature and prominent
nesting features.\cite{kosu09, kim10} These bands have a total
bandwidth of $\sim 0.3$ eV ~\cite{roth11} and derive from a
combination of the two lowest unoccupied molecular orbitals (LUMO and
LUMO+1), which have a small energy separation of $\sim$66 meV in the
molecule.\cite{kosu09}
The energy separation between the two highest occupied molecular
orbitals (HOMO and HOMO-1) is instead higher ($\sim 170$ meV) so that
the four corresponding bands in the solid are grouped in two subsets
of two bands separated by a gap of $\sim 0.1$ eV.~\cite{kosu09}
In the following, we will study within RBA the effect of doping electrons or
holes into the states immediately above (LUMO$^*$) and below (HOMO)
the gap, which are colored in blue and green, respectively, in
Fig.~\ref{fig:ele}.

\begin{figure} 
  \includegraphics[width=0.9\columnwidth]{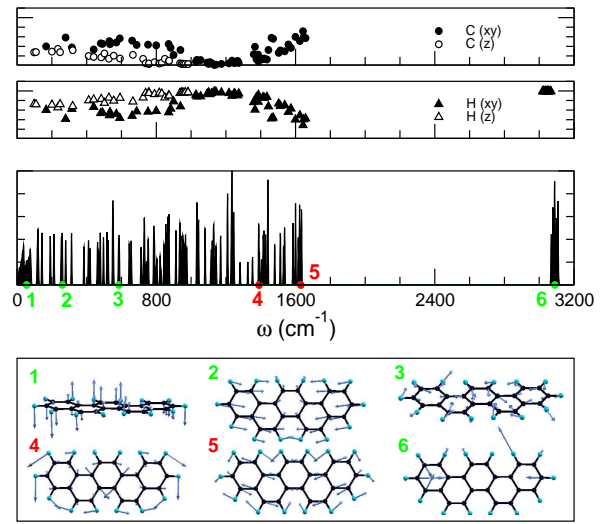}
  \caption{(Color online) Vibrational properties of picene.  {\em
      From top to bottom:} Partial C and H contribution to the phonon
    eigenvectors of a picene molecule.  Phonon DOS of (undoped) solid
    picene with numbers that mark the energy position of selected
    modes, whose $\Gamma$-point eigenvectors are shown in the bottom
    panel.  For clarity, only one picene molecule is shown.  Modes $4$
    and $5$ give the highest contribution to the $ep$ coupling, shown
    in Fig.~\ref{fig:vep}.}
  \label{fig:phon}
\end{figure}

For all dopings, we will use the phonon spectrum of undoped solid
picene, whose properties are summarized in Fig.~\ref{fig:phon}.
Most of the phonon branches of solid picene (not shown) have very
little dispersion, which reflects the molecular nature of the solid.
The dispersive intermolecular modes (acoustic, translations and
rotations of picene molecules) make up most of the phonon Density
of States (DOS) for
frequencies below $\sim$85 cm$^{-1}$.
The remaining modes are intramolecular vibration of C
and H atoms and their general character can be inferred from the top
two panels of Fig.~\ref{fig:phon}, which show the calculated partial
eigenvectors of an isolated picene molecule.
The modes below $\sim$800 cm$^{-1}$ are due to in- and out-of-plane
vibrations involving both C and H atoms that cause only small changes
in the bond length between neighbouring atoms.  Representative
examples are shown in the bottom panel of Fig.~\ref{fig:phon}. They
are out-of-plane modes that bend picene molecules with frequency of
$\sim$55 cm$^{-1}$ ({\bf 1}) and breathing modes with frequencies of
$\sim$260 ({\bf 2}) and $\sim$585 cm$^{-1}$ ({\bf 3}). At higher
frequencies, the vibrational modes cause significant change of C-C and
C-H bond lengths of adjacent atoms. The vibrations between $\sim$800
cm$^{-1}$ and $\sim$1300 cm$^{-1}$ are mostly out-of-plane, while the
modes between $\sim$1300 cm$^{-1}$ and $\sim$1600 cm$^{-1}$ have C-H
bending and C-C bending and stretching character. Two typical examples
of such modes are shown in ({\bf 4}) and ({\bf 5}), with frequencies
of $\sim$1390 and $\sim$1625 cm$^{-1}$, respectively.  The spectral
distribution of modes in this region is very similar to that of
graphene and (intercalated) graphites. The modes very high in energy
($\sim$3200 cm$^{-1}$) have in-plane C-H bond stretching character as
shown in ({\bf 6}).

\begin{figure} %% [tbp]
  \includegraphics*[width=\columnwidth]{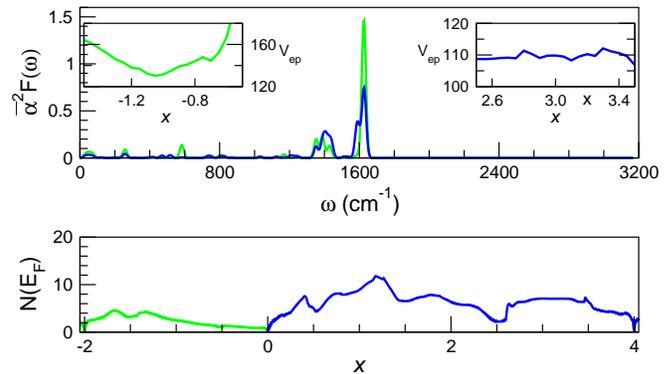}
  \caption{(Color online) {\em Upper Panel:} Spectral distribution of
    the $ep$ matrix elements of solid picene, doped with holes
    (green/light gray) or electrons (blue/dark gray).  The two insets
    show the doping dependence of the corresponding $ep$ coupling
    strength $V_{ep} = \int \omega^{-1}
    \overline{\alpha}^2F(\omega)\mathrm{d}\omega$ for different values
    of doping $x$.  {\em Lower Panel:} Electronic DOS per picene
    molecule, in states/(eV spin) as a function of doping $x$.  The
    total $ep$ coupling constant $\lambda$ as a function of doping is
    given by $\lambda(x)=N(x)V_{ep}$.  }
  \label{fig:vep}
\end{figure}

The results of our $ep$ calculations for hole and electron doped
picene are summarized in Fig.~\ref{fig:vep}. Even using the rigid band
approximation, obtaining a reliable estimate of the coupling is far
from trivial because the size of the system is large.
However, the
molecular nature of solid picene makes it possible to use
additional approximations, which made the calculations feasible.
In normal metals, the interaction between phonons and
electrons is normally expressed in terms of Eliashberg spectral
function:
%
%%%%%%%%%%%%%%%%%%%%%%%%%%%%%%%%%%%%%%%%%
%%%%%%%%%%% alpha2F %%%%%%%%%%%%%%%%%%%%%
%% sb: I corrected some obvious problems on the indices.
%% sb: Please, check that I did not screw up anything else on the way ...%
\begin{equation}
\alpha^{2}F(\omega)=\sum_{\mathbf{k},\mathbf{q},\nu,n,m}\frac{
\delta(\epsilon_{\mathbf{k}}^{n})\delta(\epsilon_{\mathbf{k+q}}^{m})}{N(E_F)}%
|g_{\mathbf{k},\mathbf{k+q}}^{\nu,n,m}|^{2}\delta(\omega-\omega
_{\nu\mathbf{q}}),%
\label{eq:alpha}
\end{equation}
%%%%%%%%%%%%%%%%%%%%%%%%%%%%%%%%%%%%%%%%%%
where $g_{\mathbf{k},\mathbf{k+q}}^{\nu,n,m}$ are the $ep$ matrix
elements that are averaged on the Fermi surface
$\delta(\epsilon_{\mathbf{k}}^{n})$.
Its first inverse moment, 
$\lambda=2\int_{0}^{\infty}d\omega\alpha^{2}F(\omega)/\omega$, 
gives the total $ep$ coupling that appears in the exponent of the
McMillan formula for the critical temperature $T_c$ of $ep$ superconductors.
Evaluating expression (~\ref{eq:alpha}) requires a careful integration
in reciprocal space over a dense set of phonon ($\mathbf{q}$) and
electron ($\mathbf{k}$) wavevectors even for elemental metals.
However, we could obtain well converged results for $\lambda$ and
$\alpha^2F(\omega)$ using affordable $\mathbf{q}$ and $\mathbf{k}$
meshes by exploiting the molecular nature of solid picene.
%o 
%o Exploiting the molecular nature of solid picene, we could obtain
%o converged results for $\lambda$ and $\alpha^2F(\omega)$ even using
%o affordable meshes of $\mathbf{q}$ and $\mathbf{k}$ points in
%o reciprocal space. (give details here)
%
First, we found the $\mathbf{q}$ dependence of $\lambda$ and
$\alpha^2F(\omega)$  to be weak, reflecting the molecular nature
of solid picene. 
Indeed, the $ep$ matrix elements $g$ are essentially intramolecular
quantities.
% because the electronic states derive from combination of
%molecular orbitals.
%
Additionally, if the Fermi surface is not pathological, the integration over
$\mathbf{k}$ and $\mathbf{q}$ in Eq.~\ref{eq:alpha} can be done
separately. The dimensionless $ep$ coupling parameter
$\lambda$ can then be factorized as $\lambda = N(E_F) V_{ep}$, where the
density of states $N(E_F)$ is determined by intermolecular
interactions and the $ep$ coupling strength $V_{ep}$ is
essentially an intramolecular quantity.\cite{lann91,devo98} Similarly,
it is possible to define a spectral function for the
$ep$ coupling strength $V_{ep}$, $\overline{\alpha}^2F(\omega)=
\alpha^2F(\omega)/N(E_F)$.
The two intramolecular quantities $\overline{\alpha}^2F(\omega)$ and
$V_{ep}$ converge much faster as a function of $\mathbf{k}$-mesh than
the intermolecular $N(E_F)$. To eliminate the noise coming from
pathological regions of the DOS, we evaluated $V_{ep}$ and
$\overline{\alpha}^2F(\omega)$ for different fillings $x$ of the
electronic bands as shown in the two insets of
Fig.~\ref{fig:vep}.\cite{foot:phon}

The values of $V_{ep}$ thus determined are
fairly constant with $x$ within a single subset of bands, as one would
expect for an intramolecular quantity. We obtain $V_{ep}=150 \pm 20$
meV for holes and $V_{ep}=110 \pm 5 $ meV for electrons. This value of
coupling is in reasonable agreement with the empirical trend $V_{ep}
=1800/N_{\pi}$, where $N_{\pi}$ is the number of atoms involved in
$\pi$ states, that has been estimated in C $\pi$ bonded molecular
systems.~\cite{devo98}

Before discussing the implications for superconductivity, it is
insightful to analyze the $V_{ep}$ spectral function
$\overline{\alpha}^2F(\omega)$, which is shown in the upper panel of
Fig.~\ref{fig:vep}.
The $ep$ spectral function has two main peaks at $\sim$1400 cm$^{-1}$
and $\sim$1600 cm$^{-1}$, which correspond to modes that bend C-H
bonds and stretch C-C bonds (modes {\bf 4} and {\bf 5} in
Fig.~\ref{fig:phon}).  Additionally, we also find smaller coupling to
modes with lower frequencies ($\omega < 800$ cm$^{-1}$).  This
spectral distribution is in agreement with the model
calculations,~\cite{kato02} which also predict a sizable contribution
at $\sim$260 cm$^{-1}$ that is present but reduced in our full-linear
response calculation.~\cite{foot:kato}

The effect of the dopants on $ep$ coupling, which is not
explicitely included in our calculations, can be
estimated from the comparison with 
graphite intercalation compounds. 
The intercalants in GICs partially reduce the $ep$ coupling
to C bond stretching modes, but also provide some additional coupling
to low-lying Einstein phonons.~\cite{mauri:GIC,boeri:GIC,kim:GIC}
These two effects balance each other to leave $\lambda$ basically
unchanged, while $\omega_{\ln}$ is reduced due to the additional
weight at low frequencies.  In contrast to GICs, which have
substantial intercalant character for states near the Fermi level, both
effects should be much weaker in solid picene because the metal states
are not expected to significantly hybridize with C $p_z$ states near
the Fermi level.~\cite{okaz10,roth11}

Accurate values of $\lambda=V_{ep}N(E_F)$ as a function of doping can
be obtained using the DOS shown in the lower panel of
Fig.~\ref{fig:vep}, which has been obtained using a (8)$^3$
$\mathbf{k}$-grid and tetrahedron method.~\cite{tetra}
For $x=3$, we get $N(E_F)$ = 7.06 states spin$^{-1}$ eV$^{-1}$,
$\lambda$ = 0.78 and $\omega_{\ln}$ = 1021 cm$^{-1}$. Using $\mu^*$ =
0.12 in the McMillan formula, we obtain a critical temperature $T_c$ =
56.5 K. This is more than three times the experimental value of $T_c$,
and we need to use a larger value of $\mu^*$ = 0.23 to reproduce the
experimental $T_c$.
In the case of hole doping, the $T_c$'s are lower due to smaller
values of DOS. For $x \sim -1.33$, with the same value of $\mu^*$ =
0.23, $N(E_F)$ = 4.31 states spin$^{-1}$ eV$^{-1}$, $\lambda$ = 0.65
and $\omega_{\ln}$ = 890 cm$^{-1}$, and we obtain a maximum $T_c$ of 6
K.

This relatively high McMillan $T_c$'s that result from strong $ep$
coupling to a few high-energy C bond-stretching phonons indicates that
superconductivity in alkali-doped picene and other hydrocarbons is
most likely phonon-mediated, and related to that of other C and
B-based superconductors such as MgB$_2$, boron-doped diamond, and
GICs.

%%% End of our results, beginning of discussion.
%Even though the $ep$ coupling in intercalated picene is large enough
%to explain the experimental $T_c$,
However, the estimates of $T_c$ based on
 Migdal-Eliashberg theory must be taken with care in the case of picene.
In fact, we have shown that the bandwidth of the conduction bands ($W \sim 300$ meV
for electrons and $\sim 1$ eV for holes), the frequencies of
strongly coupled phonons ($\omega_{ph} \sim 200$ meV), the $ep$ coupling
strength ($V_{ep} \sim$ 110-150 eV) and Coulomb repulsion ($U \sim
1.2$ eV\cite{giov10}) all have similar magnitudes.
In this range of parameters, the two most important approximations in
Migdal-Eliashberg theory of superconductivity, Migdal's theorem and
the Morel-Anderson scheme for the screening of the Coulomb repulsion,
may be invalid.
In fullerenes, whose conduction electrons have C $p_z$ character, this
regime of parameters gives rise to a variety of interesting phenomena,
the most spectacular being the occurrence of $ep$ superconductivity
near a Mott insulating phase,~\cite{gani08} well captured by
theoretical studies of local models of interacting phonons and
electrons in the non-adiabatic regime.\cite{gunn:book, mass09, baffo}

Intercalated picene is also a strong candidate for this kind of
 strongly correlated, non-adiabatic $ep$ superconductivity.
%, which is
%characterized by a bandwidth of the conduction bands that is
%comparable to phonon frequencies, $ep$ matrix elements and local
%Coulomb repulsion $U$.
%
The experimental situation is still premature to verify this claim, but 
several tests are possible for this hypothesis.
For example, Migdal-Eliashberg theory would predict a smooth behaviour of
superconductivity as a function of doping and pressure that is
governed essentially by the value of the electronic DOS at the Fermi
level.
On the other hand, strongy correlated local models would
predict a phase diagram where superconductivity exists in close
proximity to a possible Mott insulating state, which could be tuned by
doping or by a change in the intermolecular hopping driven by pressure
or intercalation of isovalent atoms with different sizes.
A very exciting prospect is the possibility of studying a wider range
of intercalated hydrocarbons by changing the number and arrangement of
benzene rings. This could be an interesting avenue for independently
tuning the $ep$ coupling~\cite{devo98} and bandwidth and degeneracy of
the conduction bands and obtaining an extensive mapping of the
parameter space for strongly correlated $ep$ superconductors.

In summary, we report the results of our first principles calculations
for the vibrational spectrum and $ep$ coupling of doped solid
picene within the rigid band approximation.
We find a large coupling of bond-stretching phonons at $\sim$1400
cm$^{-1}$ and $\sim$1600 cm$^{-1}$ both to electrons in the four
lowest conduction bands and holes in the two highest valence bands of
solid picene ($V_{ep}$ $\sim$ 110 and 150 meV, respectively). This is
sufficiently strong to explain the experimental $T_c$ of 18 K reported
for K$_3$C$_{22}$H$_{14}$ within the Migdal Eliashberg theory for $ep$
superconductors. However, due to the molecular nature of solid picene,
we argue that a strongly correlated non-adiabatic $ep$
superconductivity similar to that of alkali-doped fullerides may
occur. We propose possible experimental tests of this hypothesis and
suggest that alkali-intercalated hydrocarbons may be ideal playgrounds
to study the interplay between electronic and vibrational energy
scales in a controlled way.

We would like to thank O. Gunnarsson, R. Arita and S. Ciuchi for
useful discussion. L.B.  acknowledges hospitality of the KITP, Santa
Barbara, where part of this work was carried out.

Recently a $T_c$ of 5 K was reported in K-doped phenanthrene
(C$_{14}$H$_{10}$), ~\cite{wang11} confirming the possibility of
observing superconductivity in other polycyclic aromatic hydrocarbons.

\textit{Note added:} We have become aware of a paper by Casula
\textit{et al.}\ that studies electron-phonon coupling in
K$_3$C$_{22}$H$_{14}$.\cite{casula} They integrate over full phonon
spectrum and find $\lambda = 0.88$ and $\omega_{\ln} = 202$
cm$^{-1}$. With the caveat that our low frequency modes are less
converged, if we also integrate over the full phonon spectrum, we
obtain $\lambda = 1.41$ and $\omega_{\ln} = 240.18$ cm$^{-1}$. The
discrepancy in the value of $\lambda$ may be because our value of DOS
(7.06 states spin$^{-1}$ eV$^{-1}$ per molecule) is more than twice
larger than theirs (6.2 states spin$^{-1}$ eV$^{-1}$ per cell = 3.1
states spin$^{-1}$ eV$^{-1}$ per molecule).


\begin{references}
\bibitem{foot:calc} The results presented here are obtained using
    density functional perturbation theory\cite{dfpt} within the local
    density approximation (LDA) as implemented in the Quantum-Espresso
    package.\cite{qe} We employed norm-conserving
    pseudopotentials\cite{psps} and a plane wave basis set with a
    cut-off of 70 Ry.  Dynamical matrices were calculated on a $(2)^3$
    grid in reciprocal ($\mathbf{q}$-space), and phonon dispersions
    and DOS were then obtained by Fourier transformation of the
    real-space force constants.  For the self-consistent calculations
    of phonon frequencies, we employed a $(2)^3$ grid in
    $\mathbf{k}$-space.
\bibitem{foot:freq} Along the interlayer stacking direction
    $(0,0,1)$, two acoustic phonon modes still have small imaginary
    frequencies of -12.7 and -7.8 cm$^{-1}$ near the zone boundary.
    Convergence of these low-energy acoustic modes is a well-known
    problem also in graphite-related systems and can require very
    large plane-wave cutoffs\cite{moun05}, which are prohibitive in
    this case.  However, we find that all intramolecular modes are
    well converged with accuracies of $\sim$ 1 $\%$.
\bibitem{foot:phon} As we are focusing only on the intramolecular
    modes, we ignore the coupling of electrons to phonon frequencies
    below 100 cm$^{-1}$ when we integrate $\alpha^2F(\omega)$. If we
    integrate over the full phonon spectrum (while keeping in mind
    that our low frequency acoustic phonons are less converged), we
    get $\lambda = 1.41$ and $\omega_{\ln} = 240.18$ cm$^{-1}$. Using
    $\mu^* = 0.12$, this again gives comparable $T_c = 35.86$ K. 
    % 
    Note also that the inter- and intra-molecular phonons, which play very
    different roles in the transport properties of organic crystals,
    should also play different roles in superconductivity due to the
    different adiabatic ratio.\cite{ciuchi}.
\end{references}
\end{document}